\documentclass[final,1p,times]{elsarticle}

\usepackage{graphicx}
\usepackage{amssymb}
\usepackage{multirow}
\usepackage{natbib}

\newcommand{\mr}{\mathrm}

\newcommand{\tr}{{\mr{tr}}}

\newcommand{\be}{\begin{equation}}
\newcommand{\ee}{\end{equation}}

\begin{document}

\begin{frontmatter}

\title{Wilson mass dependence of the overlap topological charge
  density}

\author[cssm,csiro]{Peter~J.~Moran}
\ead{peter.moran@alumni.adelaide.edu.au}

\author[cssm]{Derek~B.~Leinweber}
\ead{derek.leinweber@adelaide.edu.au}

\author[cssm,china]{J.~B.~Zhang}
\ead{jbzhang08@zju.edu.cn}

\address[cssm]{Special Research Center for the Subatomic Structure of
   Matter (CSSM) and\\Department of Physics, University of Adelaide
   5005, Australia}
\address[china]{ZIMP and Department of Physics, Zhejiang University,
   Hangzhou, 310027,\\People's Republic of China}
\address[csiro]{CSIRO Mathematics, Informatics and Statistics, \\
Private Bag 33, Clayton South, VIC 3169, Australia}

\begin{abstract}
  The dependence of the overlap Dirac operator on the Wilson-mass
  regulator parameter is studied through calculations of the overlap
  topological charge densities at a variety of Wilson-mass values,
  using a L\"{u}scher-Weisz gauge action.  In this formulation, the
  Wilson-mass is used in the negative mass region and acts as a
  regulator governing the scale at which the Dirac operator is
  sensitive to topological aspects of the gauge field.  We observe a
  clear dependence on the value of the Wilson-mass and demonstrate how
  these values can be calibrated against a finite number of stout-link
  smearing sweeps.  The overlap topological charge density is also
  computed using a pre-smeared gauge field for the input kernel.  We
  show how applying the overlap operator leads to further filtering of
  the gauge field.  The results suggest that the freedom typically
  associated with smearing algorithms, through the variable number of
  sweeps, also exists in the overlap operator, through the variable
  Wilson-mass parameter.
\end{abstract}


\end{frontmatter}

\section{Introduction}
\label{sect:Introduction}

The topological structure of the QCD vacuum has been the subject of
many lattice investigations over the years.  Local patterns in
topological charge fluctuations represent a significant aspect of this
structure.  Moreover, important physical phenomena such as a large
$\eta^\prime$ mass, $\theta$ dependence, and possibly spontaneous
chiral symmetry breaking are directly related to vacuum fluctuations
of the topological charge.  By the axial anomaly, matrix elements or
correlation functions involving the topological charge density
operator $q(x)$ can be related to relevant quantities of hadronic
phenomenology.  

Lattice QCD enables non-perturbative studies of the strong interaction
from first principles, and should prove useful for studying the
important topological structure of the vacuum.  Unfortunately,
obtaining a lattice discretization for studying topology is not
completely straightforward, e.g. naively discretizing the topological
charge density generally leads to non-integer values for the
topological charge.  Physical hadronic interactions also observe an
approximate chiral symmetry that is described by the theory of QCD,
where in the massless limit, an exact chiral symmetry is realized.
Unfortunately, naive transcriptions of the continuum theory explicitly
break chiral symmetry at finite lattice spacing $a$.

The Wilson Dirac operator~\cite{Wilson:1974sk},
\begin{equation}
  D_W = \sum_\mu \left( \gamma_\mu\,\nabla_\mu - 
  \frac{1}{2}\,r\,a\,\Delta_\mu + m \right) \,,
  \label{wmass:wilsonaction}
\end{equation}
contains the irrelevant Wilson term, $r\,\Delta_\mu/2$, that
explicitly breaks chiral symmetry at $\mathcal{O}(a)$ in order to
remove fermion doublers.  This lattice discretisation is often
improved through the introduction of a clover
term~\cite{Sheikholeslami:1985ij}, however issues with chiral symmetry
breaking still exist.

One technique that has recently been used to successfully reproduce
the light hadron spectrum~\cite{Durr:2008zz}, is to filter the gauge
links prior to applying the Dirac operator.  These types of fermion
actions are typically referred to as UV-filtered or fat-link actions.
The term ``fat-link'' comes from the smeared, \emph{i.e.} fat, links
that are used to construct the Dirac operator.  One can smear either
all
links~\cite{DeGrand:1998jq,DeGrand:1998mn,Hasenfratz:2007rf,Capitani:2006ni,Durr:2008rw},
only the irrelevant
terms~\cite{Zanotti:2001yb,Zanotti:2004dr,Boinepalli:2004fz,Kamleh:2007bd},
or even just the relevant terms~\cite{Cundy:2009yy}.  Incorporating at
least some amount of UV-filtering has been shown to reduce the effects
of chiral symmetry
breaking~\cite{DeGrand:1998jq,Capitani:2006ni,DeGrand:2002vu,Boinepalli:2004fz,Kamleh:2007bd,Kamleh:2001ff}.
Unfortunately, there is no firm prescription for determining the
correct amount of smearing to apply to the gauge background.  One must
find a balance between speeding up convergence of the Dirac operator,
reducing chiral symmetry breaking effects, and removing short-distance
physics from the gauge field.  Of course, when using a fixed number of
smearing sweeps $n_{sw}$, with a constant smearing parameter $\alpha$,
the smearing procedure only introduces irrelevant terms to the action.
The fat-link action therefore remains in the same universality class
of QCD.  Nevertheless, this \emph{freedom}, in the number of smearing
sweeps that can be applied to the gauge field, can sometimes be
regarded as a drawback to fat-link fermion actions.

The difficulties with implementing exact chiral symmetry on the
lattice are summarized by the well known Nielsen-Ninomiya no-go
theorem~\cite{Nielsen:1981hk}.  The no-go theorem forbids the
existence of a local lattice Dirac operator, with exact chiral
symmetry, and is free of doublers.  However, in 1982, Ginsparg and
Wilson~\cite{Ginsparg:1981bj} showed that the physical effects of
chiral symmetry will be preserved if one can find a lattice Dirac
operator, $D$, satisfying the Ginsparg-Wilson relation,
\begin{equation}
  D \gamma_5 + \gamma_5 D = a D R \gamma_5 D \,,
  \label{wmass:GWrelation}
\end{equation}
where $R$ is a local operator.  L\"uscher later
showed~\cite{Luscher:1998pqa} that any $D$, which is a solution
of~(\ref{wmass:GWrelation}), obeys an exact chiral symmetry.  A
popular solution to the Ginsparg-Wilson relation is the Neuberger
Dirac operator~\cite{Narayanan:1994gw,Neuberger:1997fp},
\begin{equation}
  D = \frac{m}{a} \left( 1 + \frac{D_W(-m)}{\sqrt{ D_W^{\dagger}(-m)
      \,D_W(-m) }} \right) \,,
  \label{wmass:overlap_operator}
\end{equation}
which satisfies Eq.~(\ref{wmass:GWrelation}) with $R = 1/m$.  Here we
consider the standard choice of input kernel, $D_w(-m)$, the Wilson
Dirac operator with a negative Wilson-mass term.  To produce an
acceptable Dirac operator $m$ must lie in the range $0 < m < 2$.  For
$m < 0$ there are no massless fermions, while for $m > 2$ doublers
appear~\cite{Niedermayer:1998bi}.  Varying the choice of $m$ within
the allowed range results in a flow of $D$-eigenvalues, and
facilitates a scale-dependent fermionic probe of the gauge
field~\cite{Neuberger:1997fp}.  Any value of $m$ in the range $(0,2)$
should yield the same continuum
behavior~\cite{Adams:1998eg,Adams:2000rn}.  However, simulations are
performed at a finite lattice spacing $a$, and empirical studies
prefer $m \gtrsim 0.9$~\cite{Edwards:1998sh}.

The overlap Dirac operator is extremely useful for studies of QCD
vacuum structure because it satisfies the Atiyah-Singer index theorem,
and will always give an exact integer topological charge.  However,
the value is not always unique and depends on the value of the
Wilson-mass
parameter~\cite{Narayanan:1994gw,Edwards:1998sh,Narayanan:1997sa,Zhang:2001fk}.
Studies of the topological susceptibility $\chi = \langle Q^2 \rangle
/ V $, have also observed this
dependence~\cite{Edwards:1998sh,DelDebbio:2003rn}.  In particular, the
study of Ref.~\cite{DelDebbio:2003rn} found that $\chi$ varied with
$m$ for small values of $\beta$, but that this dependence decreased as
the continuum limit was approached.

In the following, we extend these previous studies to include an
analysis of the topological charge density $q(x)$, $Q \equiv \int
d^4x\ q(x)$, as $m$ is varied.  
In performing an analysis of the topological charge density, rather
than $\chi$, we have access to a greater amount of information than
that which is learnt from the susceptiblity.  A change in $\chi$ can
be due to a change in the mean-square of the topological charge
$\langle q^2(x) \rangle$, or to a more fundamental shift in the
long-range structure of the vacuum.  As such, it is not possible to
understand the underlying change in the topological structure from a
calculation of $\chi$.

A calculation of the topological charge density is also a useful probe
of the gauge field, due to its strong correlation with low-lying modes
of the Dirac operator~\cite{Kusterer:2001vk,Ilgenfritz:2008ia}, which
strongly influence how quarks propagate through the vacuum.  Also,
while our focus is on the topological charge density, all hadronic
observables on the lattice are impacted as we are examining the
properties of a lattice fermion action.  In recent years, the
available compute resources and algorithm enhancements have reached a
point where calculations of $q(x)$ using the overlap operator have
become
feasible~\cite{Ilgenfritz:2008ia,Horvath:2002yn,Ilgenfritz:2007xu}.

We visualize the topological density as this is currently the most
effective way to view the extra information.  Our analysis will focus
on a comparison between the gluonic topological charge density that is
calculated following the application of a smearing algorithm (see
Sect.~\ref{sect:dependence}).  Here our decision is motivated by the
growing relevance of fat-link fermion actions.  By studying different
smearings, we are also able to provide a direct quantitative link to
the negative-mass Wilson renormalization parameter of overlap
fermions.  We gain useful insights into the similarities and
differences between these smeared actions and the overlap action, and
their relative effectiveness for studies of topological vacuum
structure.  A central conclusion of this study, is that the
``smoothness'' of the gauge field , as seen by the overlap operator,
depends on the value of the Wilson-mass parameter

\section{Simulation details}
\label{wmass:sect:Simulationdetails}

Due to the high computational effort involved in a full calculation of
the overlap topological charge density, we consider a single slice of
representative $16^3 \times 32$ lattice configurations.  The
configurations were generated using a tadpole improved, plaquette plus
rectangle (L\"{u}scher-Weisz~\cite{Luscher:1984xn}) gauge action through the
pseudo-heat-bath algorithm, with $\beta = 4.80$ giving a lattice
spacing of $a = 0.093$~fm.

Five values of the Wilson-mass in the range $(1,2)$ are used to
calculate the overlap topological charge density,
\begin{equation}
  q_{ov}(x) = -\tr \left( \gamma_5 \left( 1 - \frac{a}{2\,m}
    D \right) \right) \,.
\end{equation}
Results are reported in terms of the input parameter $\kappa$, which
at tree level is related to $m$ by
\begin{equation}
  \kappa = \frac{1}{2\,(-m)\,a + 8\,r} \,,
\end{equation}
with the standard choice $r = 1$.  Note that the allowed range for
$\kappa$ is $1/8 < \kappa < 1/4$, and in the interacting theory
renormalization leads one to consider $1/6 \lesssim \kappa < 1/4$.  A
single calculation of $q_{ov}(x)$ for one time-slice will contain
$16^2 \times 32 = 8192$ sites of information that must be analyzed,
and this most easily achieved through direct visualizations.  In all
figures, we represent regions of positive topological charge density
by the color red fading to yellow, for large to small $q_{ov}(x)$
respectively.  Similarly, regions of negative topological charge are
colored blue fading to green.  A cutoff is applied to the topological
charge density, below which no charge is rendered.  This allows one to
observe the underlying structure of the field.

\begin{figure}
  \centering
  \begin{tabular}{c}
    \includegraphics[width=0.3\textwidth]{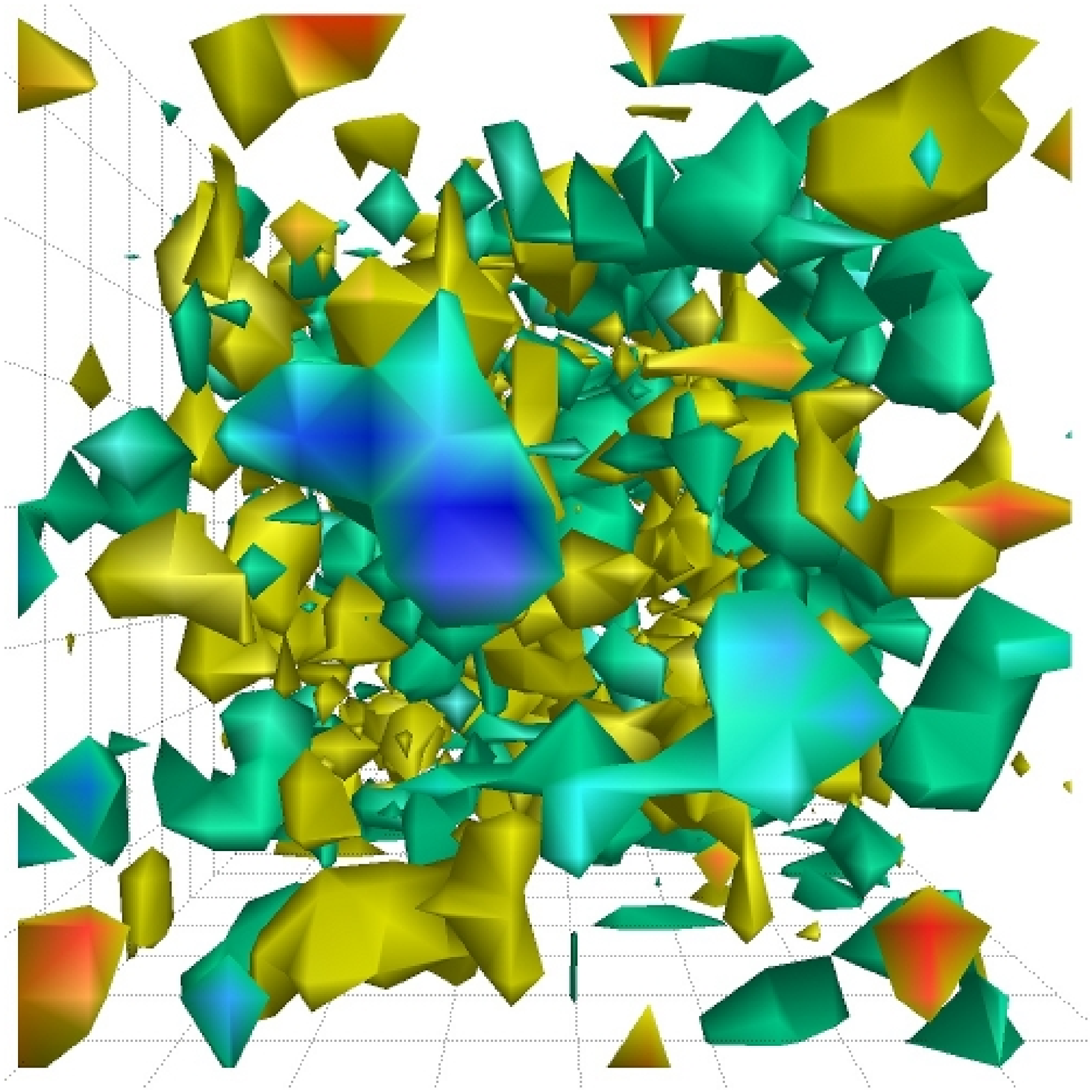} 
    \includegraphics[width=0.3\textwidth]{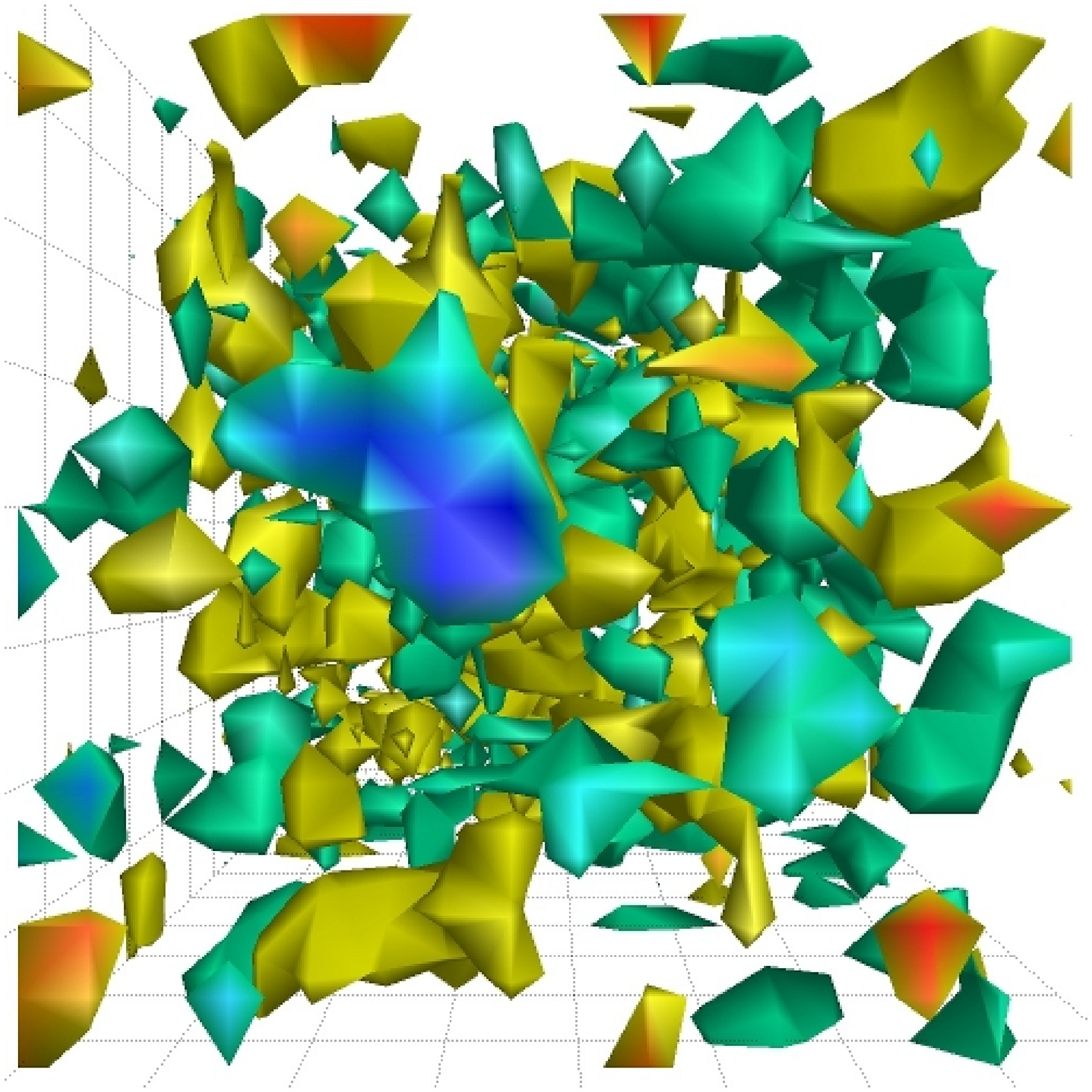} 
    \includegraphics[width=0.3\textwidth]{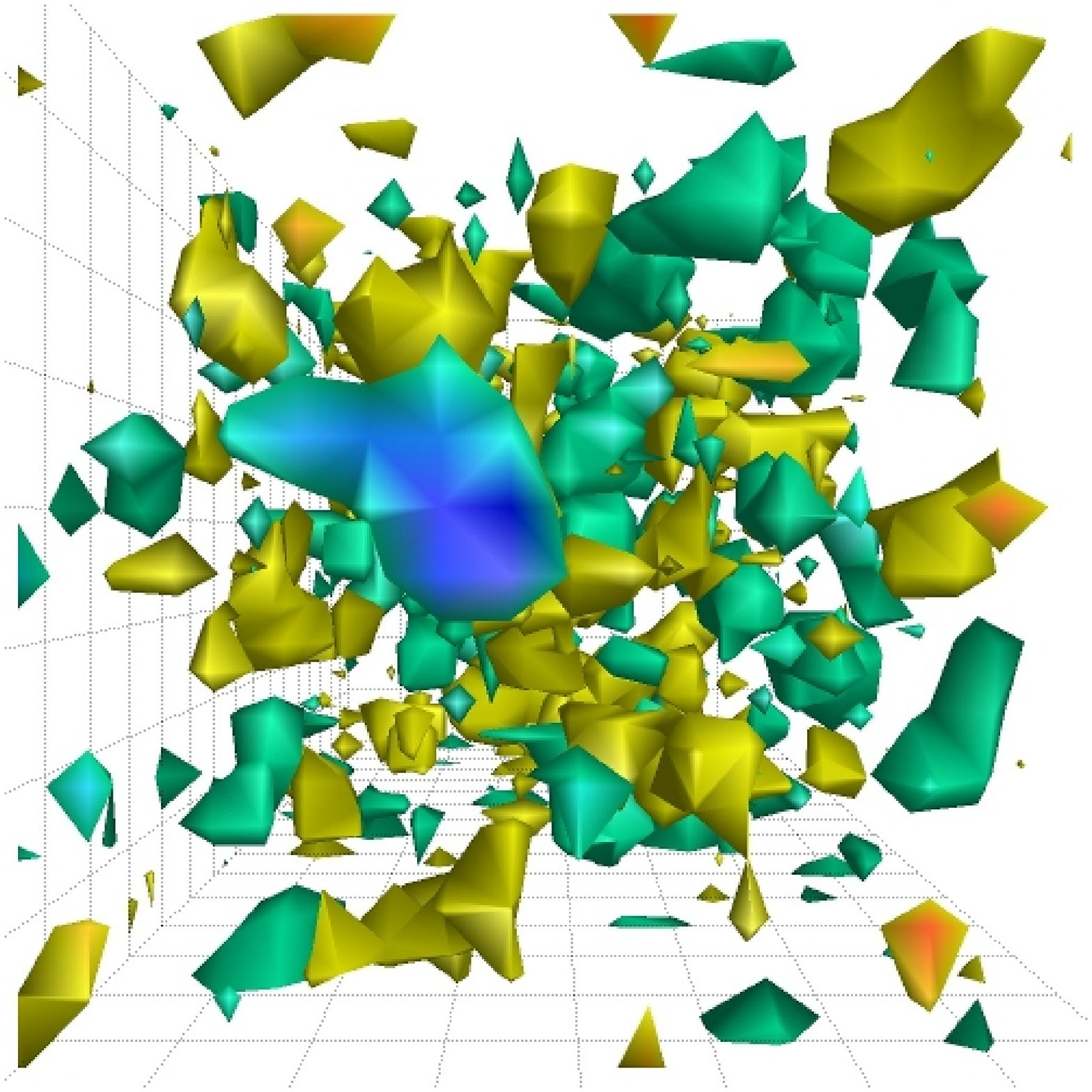} \\
    \includegraphics[width=0.3\textwidth]{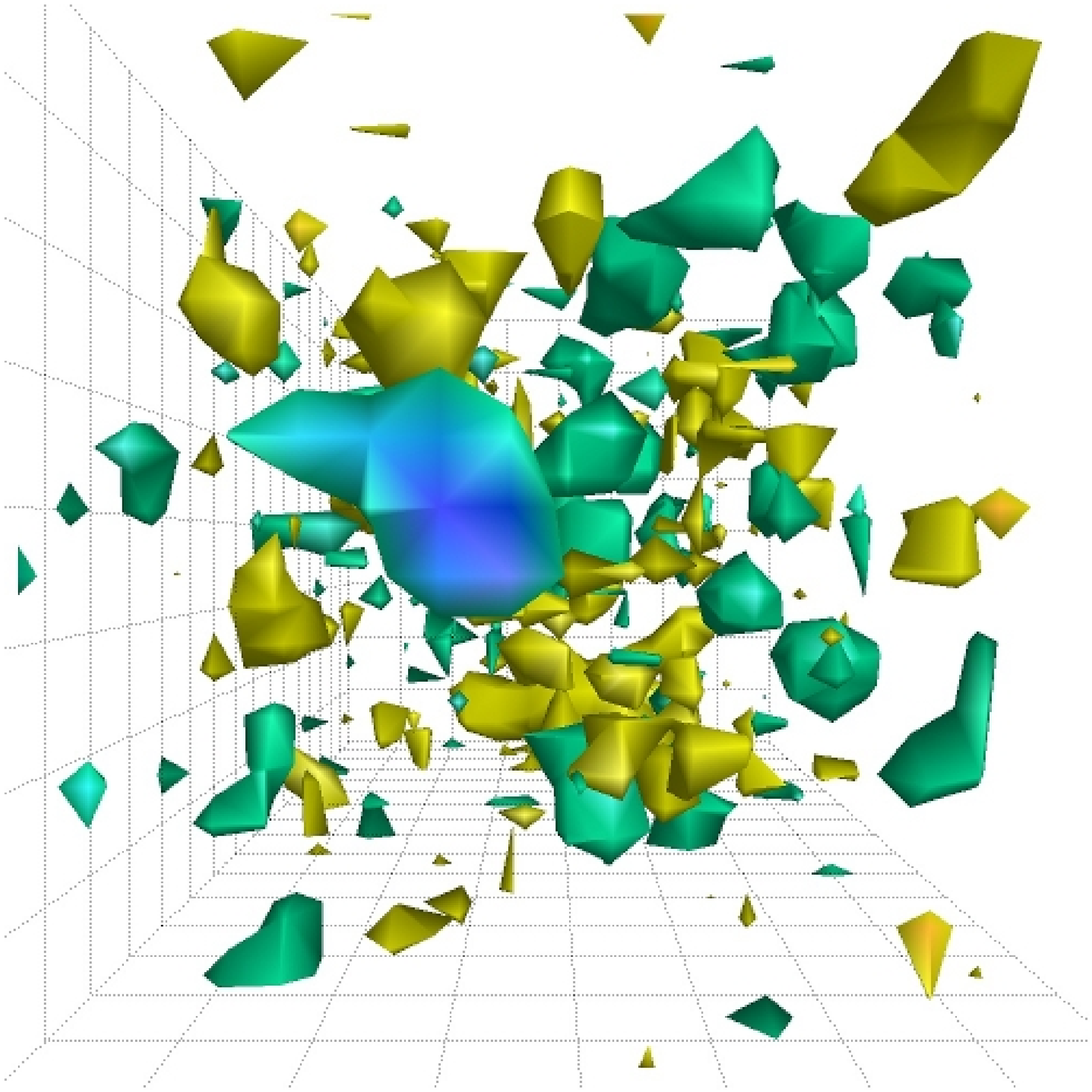}
    \includegraphics[width=0.3\textwidth]{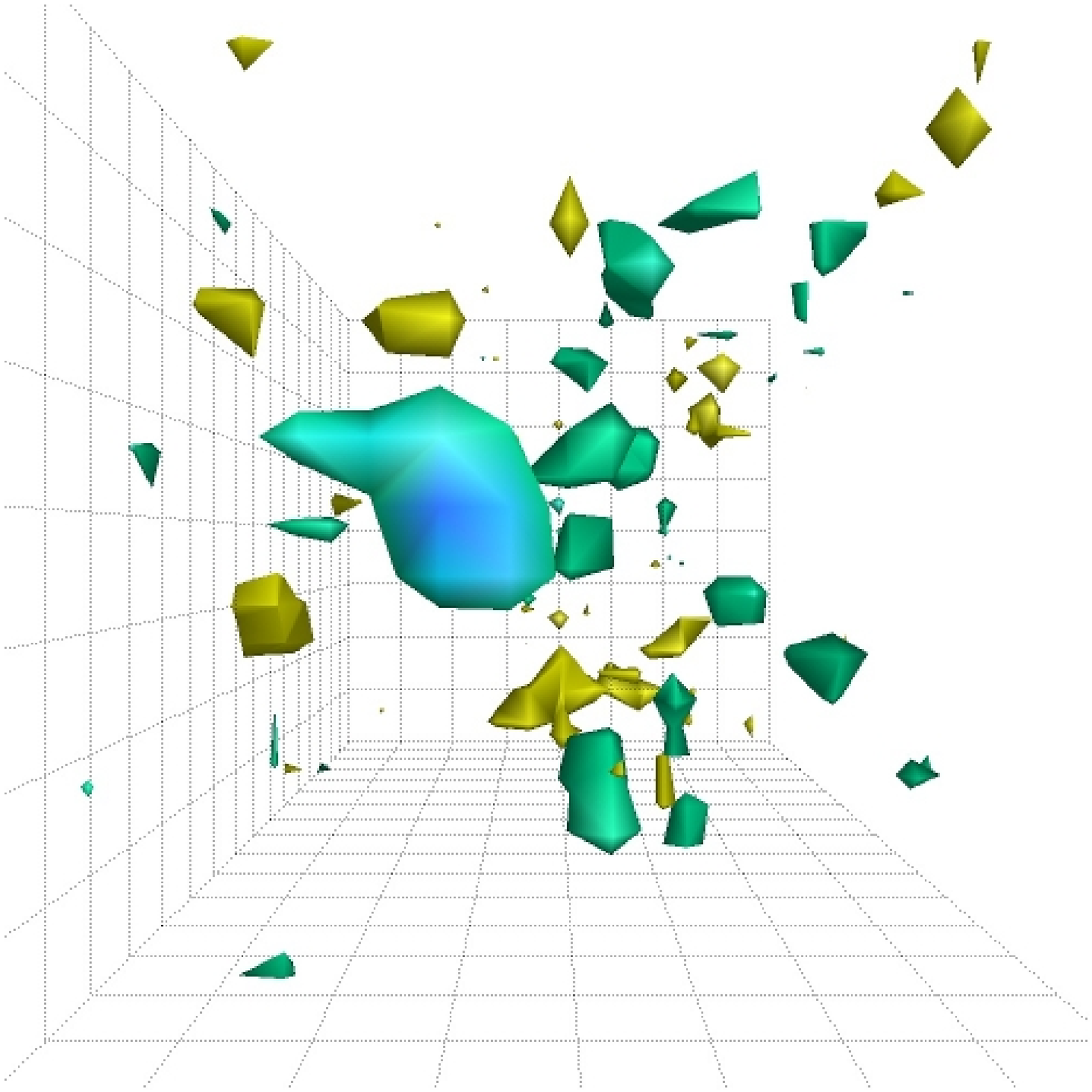} 
  \end{tabular}
  \caption{The overlap topological charge density $q_{ov}(x)$
    calculated with five choices for the Wilson hopping parameter,
    $\kappa$.  Positive regions of topological charge are colored red
    to yellow, and negative regions are shown as blue to green.  From
    left to right, we have $\kappa = 0.23$, $0.21$, and $0.19$ on the
    first row, with $0.18$, and $0.17$ on the second.  There is a
    clear dependence on the value of $\kappa$ used, with larger values
    revealing a greater amount of topological charge density.}
  \label{wmass:overlapqx}
\end{figure}

\section{Dependence on the Wilson-mass parameter}
\label{sect:dependence}

The topological charge densities, for the five choices of $\kappa$,
are presented in Fig.~\ref{wmass:overlapqx}.  A clear dependence on
$\kappa$ is apparent from the figures, with larger values of $\kappa$
revealing greater amounts of topological charge.  This is consistent
with expectations since as $\kappa$ is increased the Dirac operator
becomes more sensitive to smaller topological objects.  When using
smaller values of $\kappa$ these objects will not be felt by the Dirac
operator.

The removal of nontrivial topological objects as $\kappa$ is
decreased, bears a striking resemblance to the well tested
cooling~\cite{Berg:1981nw,Teper:1985rb,Ilgenfritz:1985dz,BilsonThompson:2002jk,BilsonThompson:2004ez,Zhang:2001fk}
and
smearing~\cite{Falcioni:1984ei,Albanese:1987ds,Bonnet:2001rc,Hasenfratz:2001hp,Morningstar:2003gk}
algorithms. In these procedures, the links on the lattice are
systematically updated such that the gauge field is driven towards a
more classical state.  This results in a removal of topological charge
density, as the action is decreased.

The over-improved stout-link smearing algorithm~\cite{Moran:2008ra} is
a modification of the original stout-link
algorithm~\cite{Morningstar:2003gk}.  Instead of the standard single
plaquette, a combination of plaquettes and rectangles are used, with
the ratio between the two tuned to preserve topology.  In every sweep
through the lattice, all links are replaced by the smeared links
$\tilde{U}_\mu(x)$~\cite{Morningstar:2003gk}
\begin{equation}
  \tilde{U}_\mu(x) = \mathrm{exp}(i Q_\mu(x) ) \, U_\mu(x) \,,
  \label{eqn:stoutlinksmearedlink}
\end{equation}
with
\begin{equation}
  Q_\mu(x) = \frac{i}{2}(\Omega_\mu^\dagger(x) - \Omega_\mu(x)) 
   - \frac{i}{6}\rm{Tr}(\Omega_\mu^\dagger(x) - \Omega_\mu(x)) \,,
\end{equation}
and
\begin{equation}
  \Omega_\mu(x) = \left( \sum_{\scriptstyle \nu \atop \scriptstyle \nu \ne \mu}
  \rho\, \Sigma_{\mu\nu}^\dagger(x) \right) U_\mu^\dagger(x)\, ,
\end{equation}
where $\Sigma_{\mu\nu}(x)$ denotes the sum of the plaquette and
rectangular staples touching $U_\mu(x)$ which reside in the $\mu-\nu$
plane.  The ratio of plaquette to rectangular staples is controlled by
a new parameter $\epsilon$~\cite{Moran:2008ra}.  In the following we
use the suggested value of $\epsilon = -0.25$, which has yielded good
results in other studies~\cite{Ilgenfritz:2008ia,Moran:2008qd}.  For
the smearing parameter we select a relatively weak value of $\rho =
0.01$.  This should be compared with the maximum value possible for
this combination of plaquettes and rectangles, $\rho \approx 0.06$.
Whilst in the standard stout-link smearing algorithm, $0.1$ is the
commonly used value.  After smearing, the gluonic topological charge
density can be calculated,
\begin{equation}
  q_{sm}(x) = \frac{g^2}{32\,\pi^2} \epsilon_{\mu\nu\rho\sigma}
  F_{\mu\nu}^{ab}(x)\,F_{\rho\sigma}^{ba}(x) \,.
\end{equation}

In order to fairly compare the two definitions for the topological
charge density one usually applies a multiplicative renormalization to
the gluonic $q_{sm}(x)$~\cite{Ilgenfritz:2008ia},
\begin{equation}
  q_{sm}(x) \rightarrow Z\,q_{sm}(x) \,.
\end{equation}
This is because after a relatively small amount of smearing the total
gluonic topological charge is typically non-integer valued due to the
presence of quantum field renormalizations.  By matching to the
overlap topological charge density we can alleviate this bias.

For this study we have a single slice of the topological charge
density and thus can not match the total topological charge.  Instead
the renormalization factor is chosen such that the \emph{structure} of
the two field densities can be best compared.  The best match to the
overlap $q_{ov}(x)$ is then found by calculating,
\begin{equation}
  \mathrm{min}\ \sum_x \left( q_{ov}(x) - Z\,q_{sm}(x) \right)^2 \,,
  \label{wmass:mindiff}
\end{equation}
as the number of smearing sweeps is varied.  Two methods for
calculating $Z$ are considered;
\begin{itemize}
\item 
  $Z_{\rm calc} \equiv \sum_x |q_{ov}(x)|\ /\ \sum_x |q_{sm}(x)| \,,$
\item
  $Z_{\rm fit}$, where the renormalization factor is 
  calculated such that~(\ref{wmass:mindiff}) is minimized.
\end{itemize}
The first definition is motived by our aim of comparing the structure
of the two field densities.  The second choice was considered to see
if the matching could be improved beyond the first definition.  We
also compare with an alternative matching
procedure~\cite{Bruckmann:2006wf,Bruckmann:2009vb} in which one
calculates,
\begin{equation}
  \Xi_{AB} = \frac{ \chi^2_{AB} }{ \chi_{AA} \, \chi_{BB} } \,,
  \label{wmass:XiAB}
\end{equation}
with
\begin{equation}
  \chi_{AB} = (1/V)\,\sum_x \left( q_A(x) - \bar{q}_A \right)
  \left( q_B(x) - \bar{q}_B \right) \,,
\end{equation}
where $\bar{q}$ denotes the mean value of $q(x)$, and in our case
$q_A(x) \equiv q_{ov(x)}$, $q_B(x) \equiv q_{sm}(x)$.  Here the best
match is found when $\Xi_{AB}$ is nearest 1.  In this case, the ratio
eliminates any dependence on the renormalization factor, $Z$.

\begin{figure}
  \begin{center}
    \begin{tabular}{cc}
      \includegraphics[width=0.35\textwidth]{c002_k23.eps} &
      \includegraphics[width=0.35\textwidth]{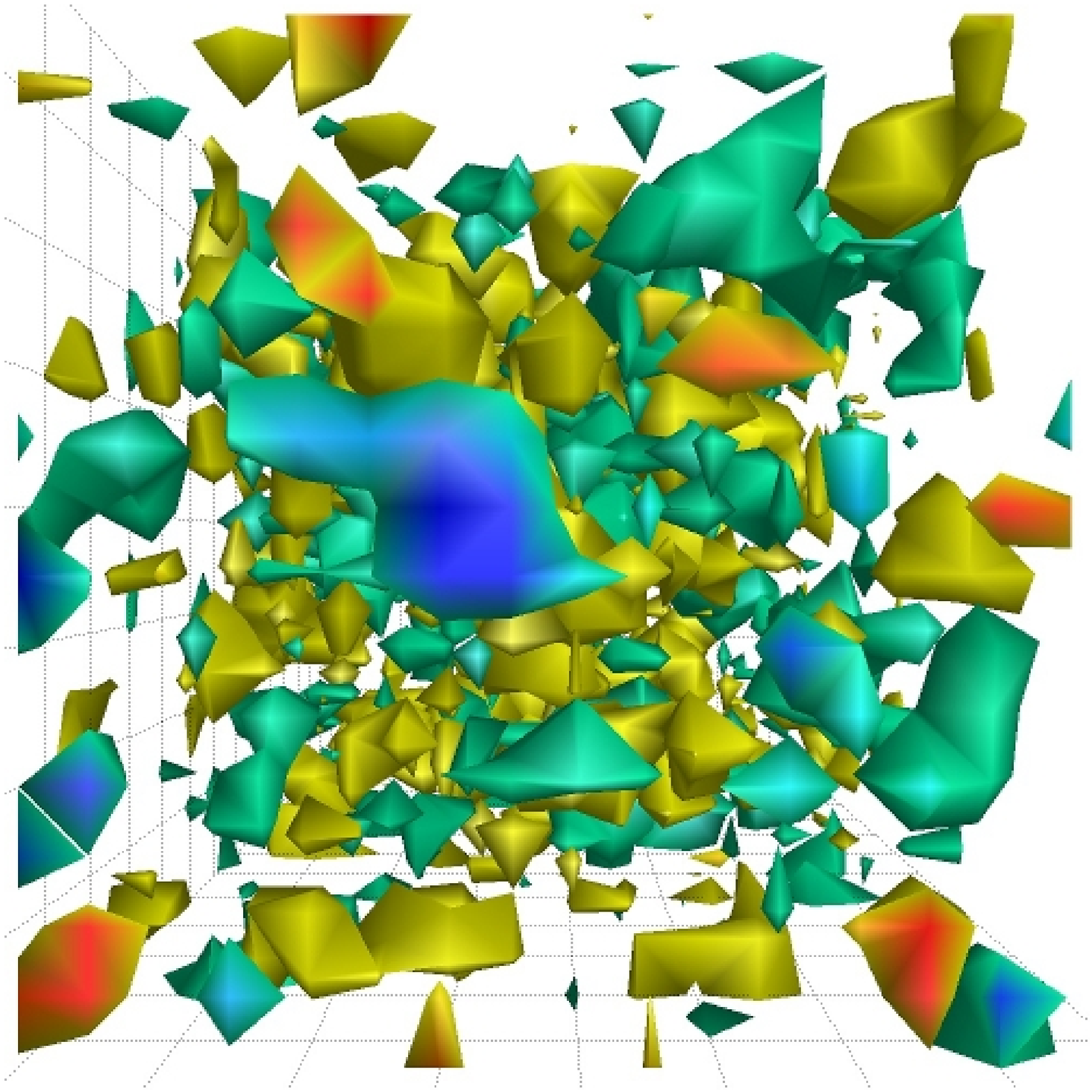} \\
      $\kappa = 0.23$ & $n_{sw} = 22$ \\
      \includegraphics[width=0.35\textwidth]{c002_k19.eps} &
      \includegraphics[width=0.35\textwidth]{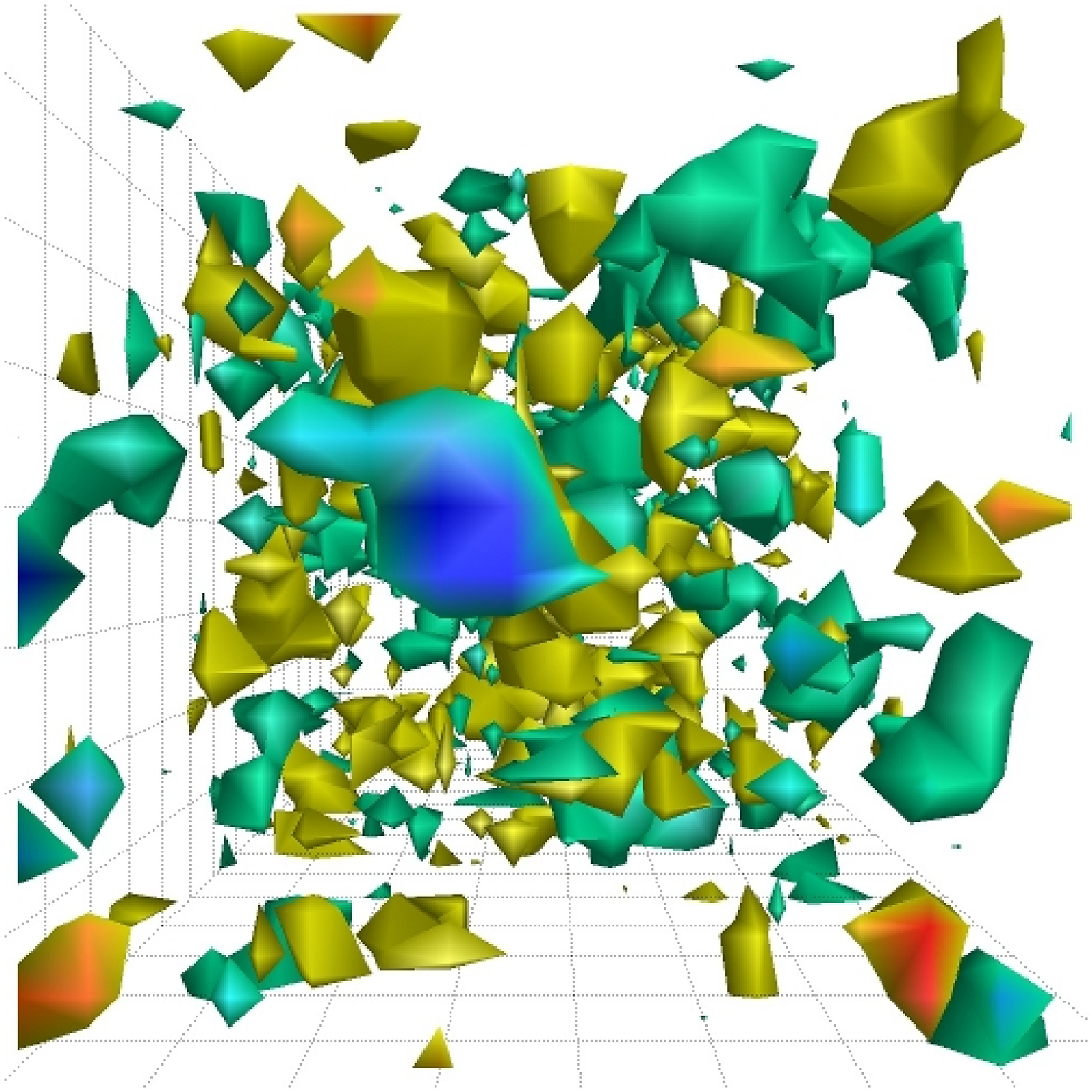} \\
      $\kappa = 0.19$ & $n_{sw} = 25$ \\
      \includegraphics[width=0.35\textwidth]{c002_k17.eps} &
      \includegraphics[width=0.35\textwidth]{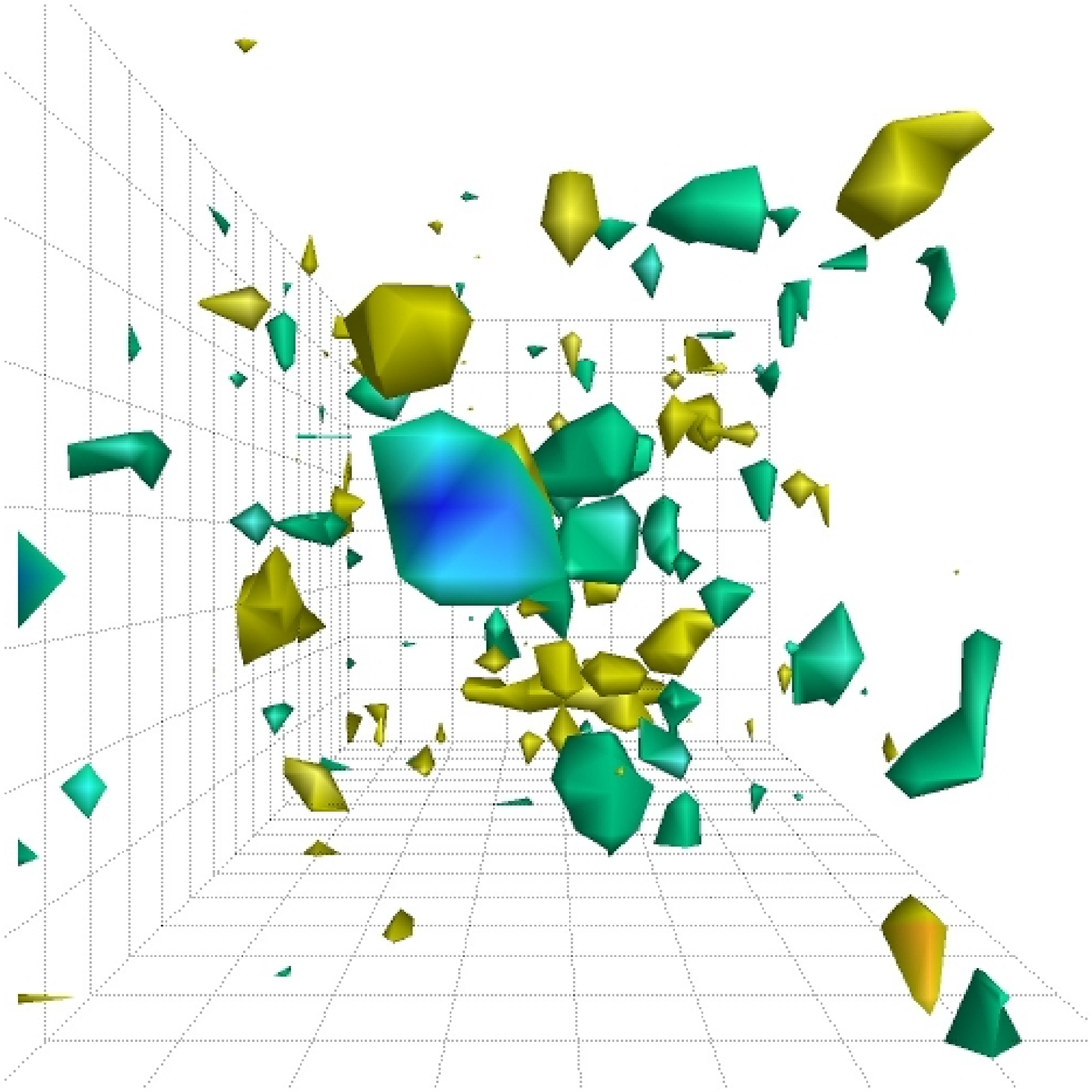} \\
      $\kappa = 0.17$ & $n_{sw} = 28$
    \end{tabular}
  \end{center}
  \caption{The best smeared matches (right) compared with the overlap
    topological charge densities (left) in order of decreasing
    $\kappa$, where $q_{sm}(x)$ is renormalised using $Z_{\rm calc}$.
    Positive regions of topological charge are colored red to yellow,
    and negative regions are shown as blue to green.  There is a clear
    relationship between $\kappa$ and $n_{sw}$, with smaller $\kappa$
    values requiring a greater number of smearing sweeps to reproduce
    the topological charge density.}
  \label{wmass:bestmatches}
\end{figure}

We first consider $Z_{\rm calc}$.  The overlap topological charge
densities, along with the corresponding best matches, for three
choices of $\kappa$ are shown in Fig.~\ref{wmass:bestmatches}.  We see
that as $\kappa$ is decreased, and non-trivial topological charge
fluctuations are removed, a greater number of smearing sweeps are
needed in order to recreate the topological charge density.  Again
this agrees with expectations since the overlap operator becomes less
sensitive to small objects as $\kappa$ is decreased, and it is these
objects that are removed by the smearing algorithm.  Comparing the
different definitions in Fig.~\ref{wmass:bestmatches} shows good
agreement in the topological structures revealed.

\begin{table}
  \begin{center}
    \begin{tabular}{|c|cc|cc|cc|}
      \hline
      $\kappa$ & $n_{sw}$ & $Z_{\rm calc}$ &
      $n_{sw}$ & $Z_{\rm fit} $ &
      $n_{sw}$ & $\Xi_{AB}$ \\
      \hline
      0.17 & 28 & 0.56 & 29 & 0.47 & 29 & 0.76 \\
      0.18 & 26 & 0.70 & 27 & 0.61 & 27 & 0.78 \\
      0.19 & 25 & 0.82 & 25 & 0.68 & 25 & 0.77 \\
      0.21 & 23 & 0.91 & 23 & 0.76 & 23 & 0.75 \\
      0.23 & 22 & 0.89 & 23 & 0.76 & 23 & 0.73 \\
      \hline
    \end{tabular}
  \end{center}
  \caption{The number of smearing sweeps, $n_{sw}$, needed to match
    the overlap topological charge density calculated with the listed
    value of $\kappa$.  The three methods used to find the best match
    are detailed in the text.}
  \label{wmass:compareZ}
\end{table}

The two methods for calculating the renormalization constant $Z$,
together with the values for $\Xi$, are compared in
Table~\ref{wmass:compareZ}.  As we move down the table there is a
monotonically increasing trend in the number of sweeps required to
match the value of $\kappa$.  We note that despite some minor
variation in $n_{sw}$, it is possible to correlate the number of
sweeps to the value of the Wilson hopping parameter.  We note that the
average renormalization factor $\bar{Z} \sim 0.7$, reflecting the fact
that with $\rho = 0.01$ the gauge fields remain rough after $\sim 25$
sweeps of smearing.  The value for $\Xi$ remains approximately
constant around $\sim 0.75$, suggesting that after renormalizing the
level of agreement between the smeared topological charge density and
the overlap density is consistent.

\section{UV-filtered overlap}

Let us now consider the effect of evaluating the overlap operator on a
pre-smeared gauge field.  This is of some relevance to UV-filtered
overlap
actions~\cite{Kamleh:2001ff,Bietenholz:2002ks,Kovacs:2002nz,Durr:2005mq},
in which all links of a gauge field are smeared prior to applying the
overlap operator.  As already seen in Fig.~\ref{wmass:bestmatches},
applying the overlap operator is in some respects similar to smearing
the gauge field.  Of interest here is whether the overlap operator,
acting on a smeared gauge field, will reveal a topological charge
density close to the input smeared gauge field, or whether further
smearing will be needed to match the calculated $q_{ov}(x)$.

\begin{figure}
  \begin{center}
    \includegraphics[width=0.45\textwidth]{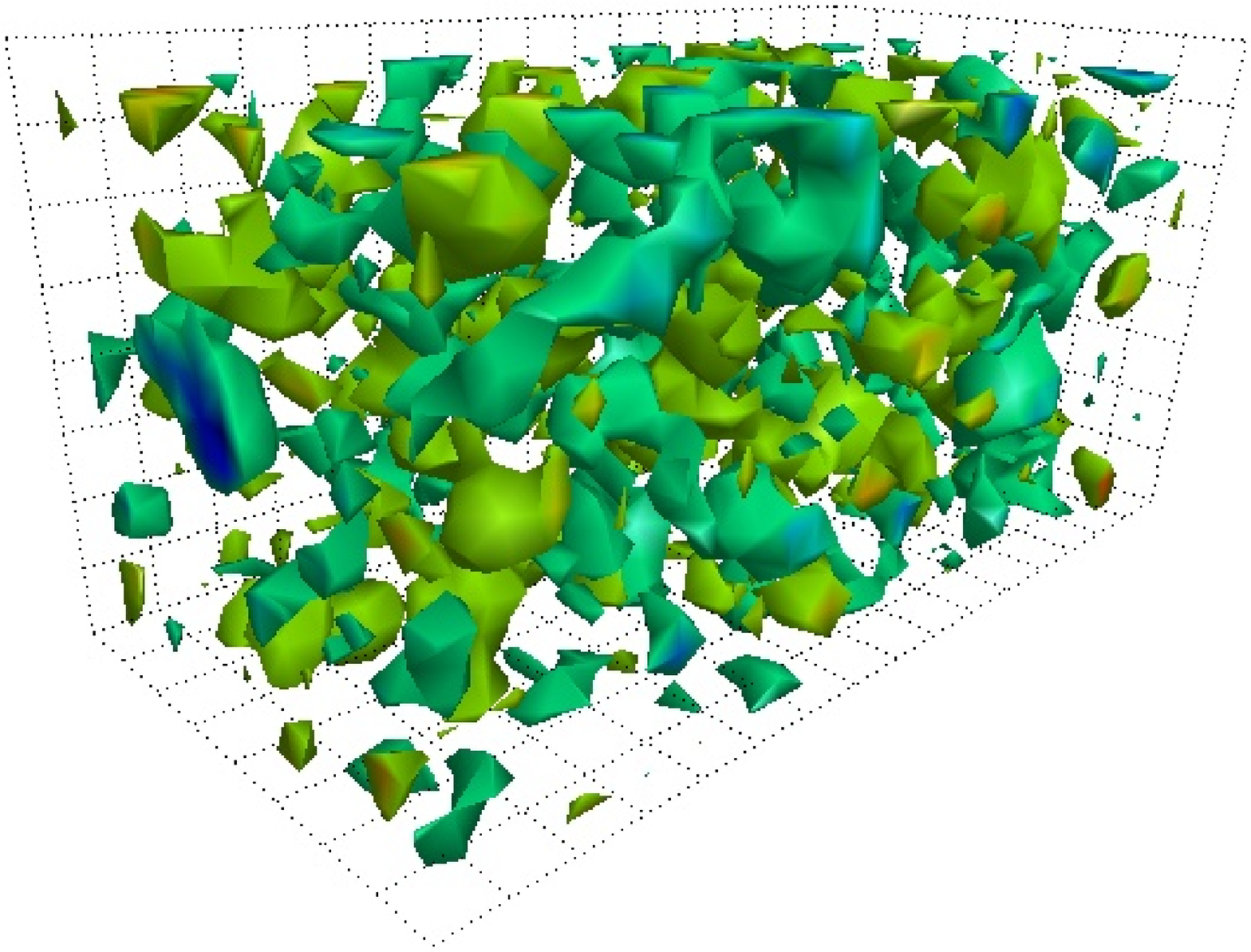}
    \includegraphics[width=0.45\textwidth]{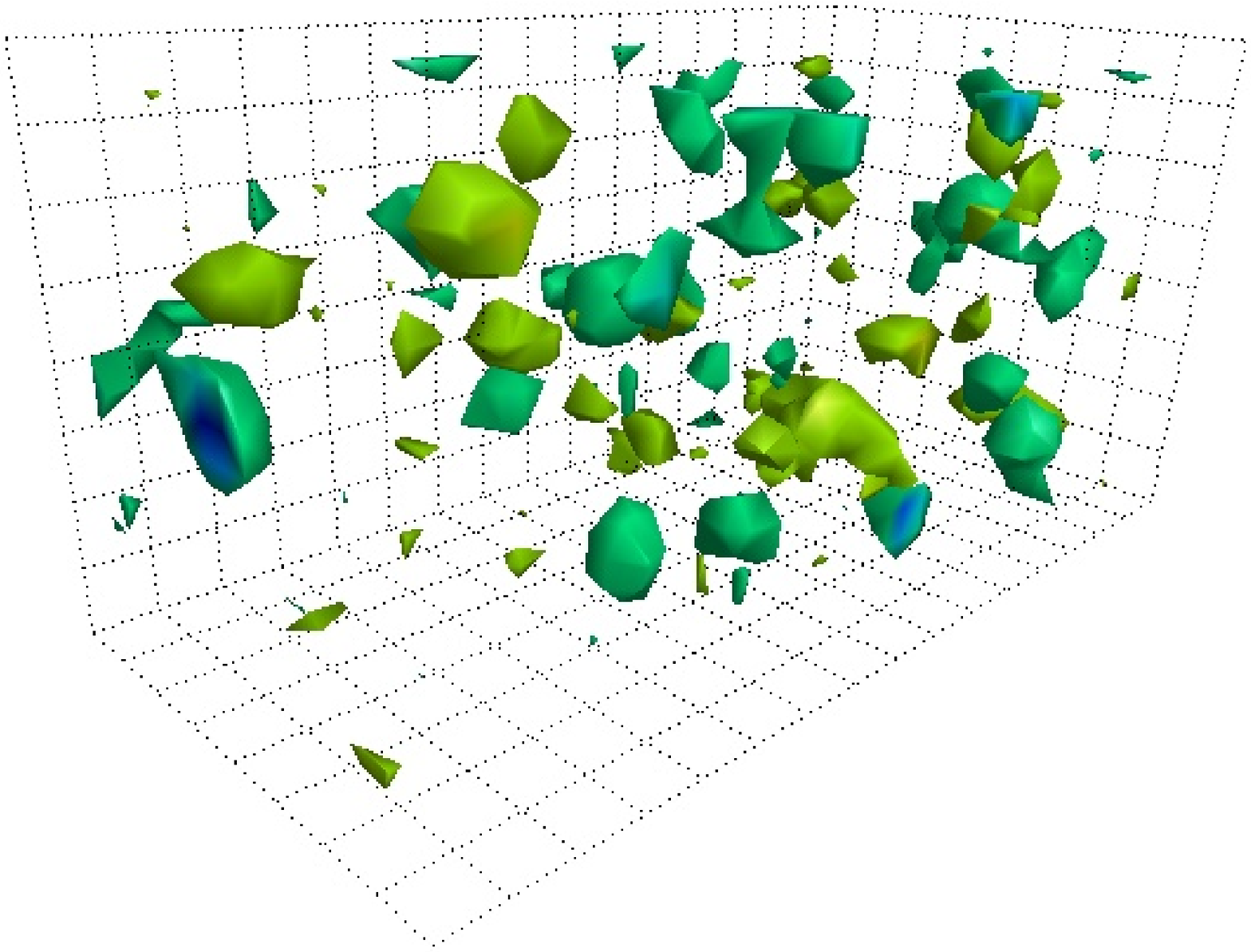}
  \end{center}
  \caption{A comparison of the overlap topological charge density
    $q_{ov}(x)$ computed using $\kappa = 0.19$ (left), with
    $q_{ov}^{\rm UV}(x)$ calculated using the same $\kappa$, on the
    same configuration, after first applying $25$ sweeps of smearing
    (right).  Positive regions of topological charge are colored
    yellow, and negative regions are shown as blue to green.}
  \label{wmass:uv_qov}
\end{figure}

To make comparisons clear, we denote the overlap topological charge
density, calculated using a smeared configuration as input, by
$q_{ov}^{\rm UV}(x)$.  We consider the third Wilson-mass, where
$\kappa = 0.19$ and the best smeared match was provided by $n_{sw} =
25$.  Figure~\ref{wmass:uv_qov} shows the original $q_{ov}(x)$ along with
the new UV-filtered $q_{ov}^{\rm UV}(x)$.  Far less topological charge
density is observed in the pre-filtered case.  Given the previous
results, it is clear that a far greater number of smearing sweeps will
be required to reproduce $q(x)$ using the gluonic definitions.

\begin{figure}
  \begin{center}
    \includegraphics[width=0.45\textwidth]{c002_ov_sw25_k19.jpg.ps}
    \includegraphics[width=0.45\textwidth]{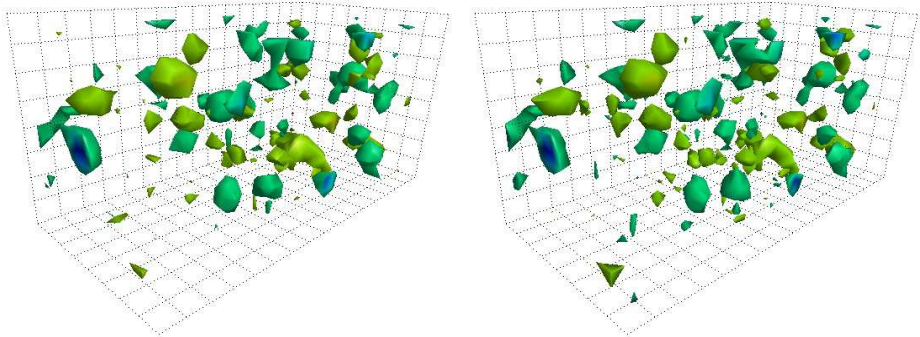}
  \end{center}
  \caption{The overlap charge density calculated on a configuration
    filtered by $25$ stout-link smearing sweeps, compared with
    $q_{sm}(x)$ after $45$ sweeps of smearing.  Positive regions of
    topological charge are colored yellow, and negative regions are
    shown as blue to green.  There is a strong correlation between the
    objects observed.  It appears as though the overlap operator has
    again ``smoothed'' the configuration.}
  \label{wmass:uv_qov_bestmatch}
\end{figure}

Repeating the same calculation as before we find that $45$ sweeps of
over-improved stout-link smearing provides the best match to the
overlap topological charge density.  A comparison between $q_{ov}^{\rm
  UV}(x)$ and the smeared $q_{sm}(x)$ is shown in
Fig.~\ref{wmass:uv_qov_bestmatch}, where $Z_\mathrm{calc} = 0.85$.
This is approximately double the original $25$ sweeps required to
match the overlap topological charge density, once again revealing the
smoothing aspect of the overlap operator. These results indicate that
the filtering that occurs in the overlap operator is independent of
the input gauge field.

\section{Conclusion}

Using direct visualizations of the topological charge density, we have
analyzed the dependence of the overlap Dirac operator on the
Wilson-mass regulator parameter $m$.  As was hinted at by previous
studies of the topological
susceptibility~\cite{Edwards:1998sh,DelDebbio:2003rn}, systematic
differences appear in the topological structure of the gauge field as
$m$ is varied.  By comparing $q_{ov}(x)$ with the gluonic definition
of the topological charge density, resolved with a topologically
stable smearing algorithm, a direct correlation between $m$ and the
number of sweeps is revealed.  Smaller values of $\kappa$ reveal
topological charge densities that are similar to using a greater
number of smearing sweeps.

From these observations, one can conclude that the ``smoothness'' of
the gauge field, as seen by the overlap operator, depends on the value
of the Wilson-mass parameter.  This is similar to fat-link fermion
actions in which the smoothness is directly dependent upon the number
of applied smearing sweeps.  These results indicate that the freedom
typically associated with fat-link fermion actions, through the number
of smearing sweeps, is also present in the overlap formalism, through
the freedom in the Wilson-mass parameter.

We also considered the application of the overlap operator to a
smeared gauge field, which is of relevance to UV-filtered overlap
actions.  We demonstrated that, regardless of the input gauge field to
the overlap operator, UV-filtering still occurs via the overlap
operator.  The strength of the filtering is of a comparable strength
to that of the overlap acting on a hot, unfiltered configuration.
When creating a UV-filtered overlap action, one must therefore take
care to preserve the short-distance physics of the gauge field.

The topological charge density revealed by the overlap operator is
similar to that revealed after 20 to 30 sweeps of stout-link smearing
with smearing parameter $\rho = 0.01$, or 2 to 3 sweeps at the
standard value of $\rho = 0.1$.  In this light, it is important to
continue investigations into the extent to which the properties and
phenomenology of the overlap operator can be obtained through the use
of an efficient Wilson-clover action on smeared configurations.

Future work could also include gauge configurations generated directly
using the overlap Dirac operator, or possibly with an alternate
overlap definition based on staggered fermions~\cite{Adams:2009eb},
which may prove more computationally efficient than the usual
Wilson-based overlap operator.

\section*{Acknowledgments}

This research was undertaken on the NCI National Facility in Canberra,
Australia, which is supported by the Australian Commonwealth
Government. We also acknowledge eResearch SA for generous grants of
supercomputing time which have enabled this project.  This research is
supported by the Australian Research Council.  J. B. Zhang is partly
supported by Chinese NSFC-Grant No.~10675101 and 10835002.


\begin{thebibliography}{10}

\bibitem{Wilson:1974sk}
Kenneth~G. Wilson.
\newblock Confinement of quarks.
\newblock {\em Phys. Rev.}, D10:2445--2459, 1974.

\bibitem{Sheikholeslami:1985ij}
B.~Sheikholeslami and R.~Wohlert.
\newblock Improved continuum limit lattice action for qcd with wilson fermions.
\newblock {\em Nucl. Phys.}, B259:572, 1985.

\bibitem{Durr:2008zz}
S.~Durr et~al.
\newblock Ab-initio determination of light hadron masses.
\newblock {\em Science}, 322:1224--1227, 2008.

\bibitem{DeGrand:1998jq}
Thomas~A. DeGrand, Anna Hasenfratz, and Tamas~G. Kovacs.
\newblock Optimizing the chiral properties of lattice fermion actions.
\newblock 1998.

\bibitem{DeGrand:1998mn}
Thomas~A. DeGrand, Anna Hasenfratz, and Tamas~G. Kovacs.
\newblock Instantons and exceptional configurations with the clover action.
\newblock {\em Nucl. Phys.}, B547:259--280, 1999.

\bibitem{Hasenfratz:2007rf}
Anna Hasenfratz, Roland Hoffmann, and Stefan Schaefer.
\newblock Hypercubic smeared links for dynamical fermions.
\newblock {\em JHEP}, 05:029, 2007.

\bibitem{Capitani:2006ni}
Stefano Capitani, Stephan Durr, and Christian Hoelbling.
\newblock Rationale for uv-filtered clover fermions.
\newblock {\em JHEP}, 11:028, 2006.

\bibitem{Durr:2008rw}
S.~Durr et~al.
\newblock Scaling study of dynamical smeared-link clover fermions.
\newblock {\em Phys. Rev.}, D79:014501, 2009.

\bibitem{Zanotti:2001yb}
James~M. Zanotti et~al.
\newblock Hadron masses from novel fat-link fermion actions.
\newblock {\em Phys. Rev.}, D65:074507, 2002.

\bibitem{Zanotti:2004dr}
J.~M. Zanotti, B.~Lasscock, D.~B. Leinweber, and A.~G. Williams.
\newblock Scaling of flic fermions.
\newblock {\em Phys. Rev.}, D71:034510, 2005.

\bibitem{Boinepalli:2004fz}
Sharada Boinepalli, Waseem Kamleh, Derek~B. Leinweber, Anthony~G. Williams, and
  James~M. Zanotti.
\newblock Improved chiral properties of flic fermions.
\newblock {\em Phys. Lett.}, B616:196--202, 2005.

\bibitem{Kamleh:2007bd}
Waseem Kamleh, Ben Lasscock, Derek~Bruce Leinweber, and Anthony~Gordon
  Williams.
\newblock Scaling analysis of flic fermion actions.
\newblock {\em Phys. Rev.}, D77:014507, 2008.

\bibitem{Cundy:2009yy}
N.~Cundy et~al.
\newblock Non-perturbative improvement of stout-smeared three flavour clover
  fermions.
\newblock {\em Phys. Rev.}, D79:094507, 2009.

\bibitem{DeGrand:2002vu}
Thomas~A. DeGrand, Anna Hasenfratz, and Tamas~G. Kovacs.
\newblock Improving the chiral properties of lattice fermions.
\newblock {\em Phys. Rev.}, D67:054501, 2003.

\bibitem{Kamleh:2001ff}
Waseem Kamleh, David~H. Adams, Derek~B. Leinweber, and Anthony~G. Williams.
\newblock Accelerated overlap fermions.
\newblock {\em Phys. Rev.}, D66:014501, 2002.

\bibitem{Nielsen:1981hk}
Holger~Bech Nielsen and M.~Ninomiya.
\newblock No go theorem for regularizing chiral fermions.
\newblock {\em Phys. Lett.}, B105:219, 1981.

\bibitem{Ginsparg:1981bj}
Paul~H. Ginsparg and Kenneth~G. Wilson.
\newblock A remnant of chiral symmetry on the lattice.
\newblock {\em Phys. Rev.}, D25:2649, 1982.

\bibitem{Luscher:1998pqa}
Martin Luscher.
\newblock Exact chiral symmetry on the lattice and the ginsparg- wilson
  relation.
\newblock {\em Phys. Lett.}, B428:342--345, 1998.

\bibitem{Narayanan:1994gw}
Rajamani Narayanan and Herbert Neuberger.
\newblock A construction of lattice chiral gauge theories.
\newblock {\em Nucl. Phys.}, B443:305--385, 1995.

\bibitem{Neuberger:1997fp}
Herbert Neuberger.
\newblock Exactly massless quarks on the lattice.
\newblock {\em Phys. Lett.}, B417:141--144, 1998.

\bibitem{Niedermayer:1998bi}
Ferenc Niedermayer.
\newblock Exact chiral symmetry, topological charge and related topics.
\newblock {\em Nucl. Phys. Proc. Suppl.}, 73:105--119, 1999.

\bibitem{Adams:1998eg}
David~H. Adams.
\newblock {Axial anomaly and topological charge in lattice gauge theory with
  overlap-Dirac operator}.
\newblock {\em Annals Phys.}, 296:131--151, 2002.

\bibitem{Adams:2000rn}
David~H. Adams.
\newblock {On the continuum limit of fermionic topological charge in lattice
  gauge theory}.
\newblock {\em J. Math. Phys.}, 42:5522--5533, 2001.

\bibitem{Edwards:1998sh}
Robert~G. Edwards, Urs~M. Heller, and Rajamani Narayanan.
\newblock Spectral flow, chiral condensate and topology in lattice qcd.
\newblock {\em Nucl. Phys.}, B535:403--422, 1998.

\bibitem{Narayanan:1997sa}
Rajamani Narayanan and Pavlos~M. Vranas.
\newblock A numerical test of the continuum index theorem on the lattice.
\newblock {\em Nucl. Phys.}, B506:373--386, 1997.

\bibitem{Zhang:2001fk}
J.~B. Zhang et~al.
\newblock {Numerical study of lattice index theorem using improved cooling and
  overlap fermions}.
\newblock {\em Phys. Rev.}, D65:074510, 2002.

\bibitem{DelDebbio:2003rn}
Luigi Del~Debbio and Claudio Pica.
\newblock Topological susceptibility from the overlap.
\newblock {\em JHEP}, 02:003, 2004.

\bibitem{Kusterer:2001vk}
Daniel-Jens Kusterer, John Hedditch, Waseem Kamleh, Derek~B. Leinweber, and
  Anthony~G. Williams.
\newblock Low-lying eigenmodes of the wilson-dirac operator and correlations
  with topological objects.
\newblock {\em Nucl. Phys.}, B628:253--269, 2002.

\bibitem{Ilgenfritz:2008ia}
E.~M. Ilgenfritz et~al.
\newblock Vacuum structure revealed by over-improved stout-link smearing
  compared with the overlap analysis for quenched qcd.
\newblock {\em Phys. Rev.}, D77:074502, 2008.

\bibitem{Horvath:2002yn}
I.~Horvath et~al.
\newblock On the local structure of topological charge fluctuations in qcd.
\newblock {\em Phys. Rev.}, D67:011501, 2003.

\bibitem{Ilgenfritz:2007xu}
E.~M. Ilgenfritz et~al.
\newblock Exploring the structure of the quenched qcd vacuum with overlap
  fermions.
\newblock {\em Phys. Rev.}, D76:034506, 2007.

\bibitem{Luscher:1984xn}
M.~Luscher and P.~Weisz.
\newblock On-shell improved lattice gauge theories.
\newblock {\em Commun. Math. Phys.}, 97:59, 1985.

\bibitem{Berg:1981nw}
B.~Berg.
\newblock Dislocations and topological background in the lattice o(3) sigma
  model.
\newblock {\em Phys. Lett.}, B104:475, 1981.

\bibitem{Teper:1985rb}
M.~Teper.
\newblock Instantons in the quantized su(2) vacuum: A lattice monte carlo
  investigation.
\newblock {\em Phys. Lett.}, B162:357, 1985.

\bibitem{Ilgenfritz:1985dz}
Ernst-Michael Ilgenfritz, M.~L. Laursen, G.~Schierholz, M.~Muller-Preussker,
  and H.~Schiller.
\newblock First evidence for the existence of instantons in the quantized su(2)
  lattice vacuum.
\newblock {\em Nucl. Phys.}, B268:693, 1986.

\bibitem{BilsonThompson:2002jk}
Sundance~O. Bilson-Thompson, Derek~B. Leinweber, and Anthony~G. Williams.
\newblock Highly-improved lattice field-strength tensor.
\newblock {\em Ann. Phys.}, 304:1--21, 2003.

\bibitem{BilsonThompson:2004ez}
S.~O. Bilson-Thompson, D.~B. Leinweber, A.~G. Williams, and G.~V. Dunne.
\newblock {Comparison of $|$Q$|$ = 1 and $|$Q$|$ = 2 gauge-field configurations
  on the lattice four-torus}.
\newblock {\em Annals Phys.}, 311:267--287, 2004.

\bibitem{Falcioni:1984ei}
M.~Falcioni, M.~L. Paciello, G.~Parisi, and B.~Taglienti.
\newblock Again on su(3) glueball mass.
\newblock {\em Nucl. Phys.}, B251:624--632, 1985.

\bibitem{Albanese:1987ds}
M.~Albanese et~al.
\newblock Glueball masses and string tension in lattice qcd.
\newblock {\em Phys. Lett.}, B192:163--169, 1987.

\bibitem{Bonnet:2001rc}
Frederic D.~R. Bonnet, Derek~B. Leinweber, Anthony~G. Williams, and James~M.
  Zanotti.
\newblock Improved smoothing algorithms for lattice gauge theory.
\newblock {\em Phys. Rev.}, D65:114510, 2002.

\bibitem{Hasenfratz:2001hp}
Anna Hasenfratz and Francesco Knechtli.
\newblock Flavor symmetry and the static potential with hypercubic blocking.
\newblock {\em Phys. Rev.}, D64:034504, 2001.

\bibitem{Morningstar:2003gk}
Colin Morningstar and Mike~J. Peardon.
\newblock Analytic smearing of su(3) link variables in lattice qcd.
\newblock {\em Phys. Rev.}, D69:054501, 2004.

\bibitem{Moran:2008ra}
Peter~J. Moran and Derek~B. Leinweber.
\newblock Over-improved stout-link smearing.
\newblock {\em Phys. Rev.}, D77:094501, 2008.

\bibitem{Moran:2008qd}
Peter~J. Moran and Derek~B. Leinweber.
\newblock Impact of dynamical fermions on qcd vacuum structure.
\newblock {\em Phys. Rev.}, D78:054506, 2008.

\bibitem{Bruckmann:2006wf}
Falk Bruckmann et~al.
\newblock Quantitative comparison of filtering methods in lattice qcd.
\newblock {\em Eur. Phys. J.}, A33:333--338, 2007.

\bibitem{Bruckmann:2009vb}
F.~Bruckmann et~al.
\newblock Comparison of filtering methods in su(3) lattice gauge theory.
\newblock {\em PoS}, CONFINEMENT8:045, 2008.

\bibitem{Bietenholz:2002ks}
Wolfgang Bietenholz.
\newblock Convergence rate and locality of improved overlap fermions.
\newblock {\em Nucl. Phys.}, B644:223--247, 2002.

\bibitem{Kovacs:2002nz}
Tamas~G. Kovacs.
\newblock Locality and topology with fat link overlap actions.
\newblock {\em Phys. Rev.}, D67:094501, 2003.

\bibitem{Durr:2005mq}
Stephan Durr, Christian Hoelbling, and Urs Wenger.
\newblock Physics prospects of uv-filtered overlap quarks.
\newblock {\em Nucl. Phys. Proc. Suppl.}, 153:82--89, 2006.

\bibitem{Adams:2009eb}
David~H. Adams.
\newblock {Theoretical foundation for the Index Theorem on the lattice with
  staggered fermions}.
\newblock {\em Phys. Rev. Lett.}, 104:141602, 2010.

\end{thebibliography}




\end{document}